\documentclass[aps,prl,twocolumn, reprint, amsmath, amssymb]{revtex4-1}
\usepackage[centering,hmargin=2cm,vmargin=2cm]{geometry}
\newcommand{\const}{\,{\rm const}\,}
\newcommand{\td}{\text{d}}
\def\be{\begin{equation}}
\def\ee{\end{equation}}
\def\bea{\begin{eqnarray}}
\def\eea{\end{eqnarray}}

\begin{document}

\title{A supersymmetric black lens}

\author{Hari K Kunduri}
\email{hkkunduri@mun.ca}
\affiliation{Department of Mathematics and Statistics, Memorial University of Newfoundland, St John's, Canada}

\author{James Lucietti}
\email{j.lucietti@ed.ac.uk}
\affiliation{School of Mathematics and Maxwell Institute of Mathematical Sciences, University of Edinburgh, King's Buildings, Edinburgh, UK}

\begin{abstract}
We present a new supersymmetric, asymptotically flat,  black hole solution to five-dimensional supergravity. It is regular on and outside an event horizon of lens space topology $L(2,1)$. It is the first example of an asymptotically flat black hole with lens space topology. The solution is characterised by a charge, two angular momenta and a magnetic flux through a non-contractible disc region ending on the horizon, with one constraint relating these.
\end{abstract}

\maketitle

A fundamental result in the theory of black holes is Hawking's horizon topology theorem~\cite{Hawking:1971vc}. It shows that for asymptotically flat, stationary black holes satisfying the dominant energy condition, cross-sections of the event horizon must be topologically $S^2$.  
It has been known for over a decade that black holes in higher dimensions are not so constrained. In five dimensions, an explicit example of an asymptotically flat black hole with horizon topology $S^1\times S^2$ -- a black ring -- was presented~\cite{Emparan:2001wn}. In conjunction with the $S^3$ topology Myers-Perry black hole, this explicitly demonstrated black hole non-uniqueness in five dimensional vacuum gravity~\cite{Emparan:2008eg}.

Hawking's horizon topology theorem was subsequently generalised to higher dimensions, revealing a weaker constraint on the topology, namely, cross-sections of the horizon must have a positive Yamabe invariant~\cite{Galloway:2005mf}. However, it is unclear whether every topology allowed by this theorem is actually realised by a black hole solution. So far, the black ring is the only non-spherical example known  with a connected horizon, although it is believed many other types exist \cite{Emparan:2009vd, Kunduri:2010vg}.

In five dimensions, the positive Yamabe condition allows for $S^3, S^1\times S^2$, quotients of $S^3$ by a discrete subgroup, and connected sums of these.  In the context of stationary solutions with $U(1)^2$ rotational symmetry, it has been shown that the possible topologies are further constrained to be one of $S^3, S^1\times S^2$, or $L(p,q)$ where $L(p,q) \cong S^3 /\mathbb{Z}_p$ is a lens space~\cite{Hollands:2007aj}. The former two topologies are of course already realised by the Myers-Perry solutions and black rings. There have been various attempts at finding an asymptotically flat black hole solution with lens space horizon topology -- a black lens --  to the vacuum Einstein equations, although they have all resulted in solutions with naked singularities~\cite{Evslin:2008gx, Chen:2008fa}.

In this note we show that black lenses do in fact exist, by writing down a simple supersymmetric, asymptotically flat, black lens solution to five dimensional minimal supergravity. Specifically, we construct an example that is regular on and outside an event horizon with lens space topology $L(2,1) \cong \mathbb{RP}^3 \cong S^3/\mathbb{Z}_2$.

The bosonic content of five-dimensional minimal supergravity is a metric $g$ and a Maxwell field $F$. The general form for supersymmetric solution was found in~\cite{Gauntlett:2002nw},
\bea
\td s^2 = -f^2( \td t + \omega)^2 + f^{-1} \td s^2_M \; ,
\eea
where $V = \partial /\partial t$ is the supersymmetric Killing vector field, $\td s^2_M$ is a hyper-K\"ahler base and $f, \omega$ are a function and 1-form on the base $M$. We will choose the base to be a Gibbons-Hawking space,
\begin{equation}
\td s^2_M = H^{-1} (\td \psi + \chi_i \td x^i)^2 + H \td x^i \td x^i \; ,
\end{equation} 
where $x^i, i=1,2,3$, are Cartesian coordinates on $\mathbb{R}^3$, the function $H$ is harmonic on $\mathbb{R}^3$ and $\chi$ is a 1-form on $\mathbb{R}^3$ satisfying $\star_3 \td  \chi = \td H$.  As is well known~\cite{Gauntlett:2002nw}, such solutions are then specified by 4 harmonic functions $H,K,L,M$, in terms of which,
\bea
 f^{-1} &=& H^{-1} K^2 + L \; ,\\
 \omega &=& \omega_\psi(\td \psi + \chi_i \td x^i) + \hat{\omega}_i \td x^i \; , \nonumber
 \eea
 where,
 \begin{eqnarray}
 \label{omega}
 \omega_\psi &=& H^{-2} K^3 + \tfrac{3}{2}H^{-1}KL + M \; , \nonumber  \\  \star_3 \td \hat\omega &=& H \td M - M \td H + \tfrac{3}{2}(K \td L - L \td K)
   \; .
 \end{eqnarray} 
 The Maxwell field is determined by
 \be
 \label{max}
 F = \tfrac{\sqrt{3}}{2} \td \left[ f( \td t + \omega) - \frac{K}{H} (\td \psi+ \chi_i \td x^i)  - \xi_i \td x^i\right]  \; ,
 \ee
where the 1-form $\xi$ satisfies  $\star_3 \td \xi = - \td K$.

Now we write the $\mathbb{R}^3$ in polar coordinates,
\be
\td x^i \td x^i = \td r^2 + r^2 (\td\theta^2 + \sin^2\theta \td\phi^2) \; ,
\ee
and consider the 2-centred solution 
\begin{eqnarray}
H &=& \frac{2}{r} - \frac{1}{r_1} \; , \qquad M = m + \frac{m_1}{r_1} \; , \nonumber  \\
K &=& \frac{k_0}{r} + \frac{k_1}{r_1} \; ,\qquad L = 1 + \frac{\ell_0}{r} + \frac{\ell_1}{r_1} \; ,
\end{eqnarray}  where $r_1 = \sqrt{r^2 + a_1^2 - 2ra_{1} \cos\theta}$ is the distance from the origin to the `centre' $(0,0,a_1)$. We assume $a_1>0$.  We used a shift freedom in the harmonic functions to remove any $1/r$ term in $M$, without any loss of generality~\cite{Bena:2005va}.  To fully determine the solution, we must integrate to find the 1-forms $\chi, \hat{\omega}, \xi$.  We find,
\bea
\chi &=& \left[ 2\cos\theta - \frac{r\cos\theta - a_1}{r_1} \right] d\phi  \;, \nonumber  \\
\hat{\omega}  &=& \left[ - ( 2m + \tfrac{3}{2} k_0) \cos \theta + \frac{(m- \tfrac{3}{2} k_1)(r\cos\theta -a_1)}{r_1} \right. \nonumber \\ &+& \left. \frac{\left( r-a_1\cos \theta \right) (2m_1+\tfrac{3}{2}( \ell_1 k_0 - \ell_0 k_1))}{a_1 r_1} +c \right] \td \phi  \; ,\nonumber \\
\xi &=& - \left[ k_0 \cos\theta + k_1 \frac{r\cos\theta - a_1}{r_1} + c'\right] d\phi   \; ,
\eea
where $c,c'$ are integration constants (we have set the one for $\chi$ to zero by suitably shifting $\psi$). Crucially, observe that $\chi \sim \cos \theta \td \phi$ as $r \to \infty$, and $\chi \sim (1+2\cos \theta ) \td \phi$ as $r\to 0$; as we will show this allows the spacetime to interpolate between $S^3$ at spatial infinity and $S^3/\mathbb{Z}_2$ near the horizon.

For a suitable choice of constants, the solution is asymptotically flat. Defining $r = \rho^2/4$, it is easy to check that the Gibbons-Hawking base  for $\rho \to \infty$ looks like
 \bea
 \td s^2_M &\sim& \td \rho^2 + \tfrac{1}{4}\rho^2 (\td \theta^2 + \sin^2 \theta \td \phi^2 ) \nonumber \\ &+& \tfrac{1}{4}\rho^2( \td \psi + \cos \theta \td \phi )^2 \; , \label{R4}
 \eea
with subleading terms of order $\mathcal{O}(\rho^{-2})$. Hence the base is asymptotically $\mathbb{R}^4$ provided we fix the periods of the angles to be $\Delta \psi = 4\pi$, $\Delta \phi = 2\pi$ and $0 \leq \theta \leq \pi$. Now, it is also clear that $ f = 1+\mathcal{O}(\rho{^{-2}})$.  Furthermore, $\omega_\psi = \mathcal{O}(\rho^{-2})$ and $\omega_\phi = \mathcal{O}(\rho^{-2})$, provided we fix the constants,
\bea
m =-\tfrac{3}{2} (k_0+k_1) \; ,  \; \;
c = \frac{3 \ell_0 k_1  - 3 \ell_1 k_0 -4m_1}{2a_1} ,
\eea
respectively.  We will assume these choices henceforth, so our solution is asymptotically flat $\mathbb{R}^{1,4}$.

Although the solution appears singular at `centres' $r=0$ and $r_1=0$, we will show that by suitably choosing our constants, $r=0$ corresponds to an event horizon and $r_1=0$ corresponds to a smooth timelike point.

First consider the centre $r_1=0$. Near this centre the Gibbons-Hawking base approaches $-\mathbb{R}^4$ smoothly, provided the angles are identified in the same manner as required by asymptotic flatness~\cite{Kunduri:2014iga, Bena:2005va}. To see this, change to $\mathbb{R}^3$ polar coordinates with respect to this centre, $(r_1,\theta_1)$, then set $\rho= 2 \sqrt{r_1}$.  One finds that $\td s^2_M$ as $\rho \to 0$ approaches minus (\ref{R4}) with $(\theta,\psi, \phi)$ replaced by $(\theta_1,\psi+2\phi, -\phi)$. Introducing $\mathbb{R}^2$ polar coordinates $(X,\Phi)$, $(Y, \Psi)$ on the orthogonal 2-planes, $X=\rho \cos \tfrac{1}{2} \theta_1, \; Y= \rho \sin \tfrac{1}{2} \theta_1$, $\Phi = \tfrac{1}{2} (\psi+ \phi)$ and $\Psi= \tfrac{1}{2} ( \psi+ 3 \phi)$, one can then demonstrate smoothness at the centre~\cite{Kunduri:2014iga}. 
Further, imposing that the centre is a timelike point requires $f|_{{\bf x} = {\bf x}_1} \neq 0$, which implies $\ell_1= k_1^2$. In fact, to get the correct spacetime signature we require $f|_{{\bf x} = {\bf x}_1}<0$.   One can then check the function $f$ is smooth at the centre $r_1 =0$. Since $\partial_\psi$ degenerates at the centre, smoothness also requires that $V \cdot \partial_\psi = -f^2 \omega_\psi$ vanishes at that point. In fact, $\omega_\psi$ is singular at the centre unless $m_1 = \tfrac{1}{2}k_1^3$.
Then further imposing $\omega_\psi$ vanishes at the centre also implies the constraint
\be
3a_1(2k_1 + k_0) + 3\ell_0 k_1 - 3k_0 k_1^2 - 2k_1^3 = 0  \; .  \label{constr}
\ee
These conditions imply $\omega = \mathcal{O}(X^2) \td \Phi + \mathcal{O}(Y^2) \td \Psi$ ensuring the 1-form $\omega$ - and hence the spacetime metric - is smooth at the centre $r_1=0$.  The Maxwell field is then also smooth at the centre. Thus our solutions are parameterised by $(\ell_0, k_0, k_1, a_1)$, subject to the constraint (\ref{constr}), resulting in a 3-parameter family. Observe that if $k_1 = 0$, then $k_0=0$; we will show this is incompatible with smoothness  of the axis of rotation (see (\ref{ineq1},\ref{ineq2})). Thus $k_1 \neq 0$ which allows us to solve (\ref{constr}) for $\ell_0$.

Now consider the centre $r=0$.  We will show that this corresponds to a regular event horizon if
\be
R_1^2 \equiv 2\ell_0 + k_0^2>0 \;  , \quad R_2^2 \equiv  \frac{\ell_0^2(8 \ell_0 + 3 k_0^2)}{ (2\ell_0 + k_0^2)^2}>0  \; . \label{regularhorizon}
\ee
To this end, transform to new coordinates $(v,r,\psi', \theta, \phi)$,
\bea
\td t &=& \td v + \left( \frac{A_0}{r^2} + \frac{A_1}{r} \right) \td r  \; , \nonumber \\  \td \psi + \td \phi &=& d\psi' + \frac{B_0}{r} \td r \; ,
\eea
where $A_0, A_1, B_0$ are constants to be determined. Then,
\bea
g_{vv} &=& - \frac{4r^2}{R_1^4} + \mathcal{O}(r^3) \; , \quad g_{\psi'\psi'} = \tfrac{1}{4} R_2^2 + \mathcal{O}(r) \; ,\nonumber \\
g_{v \psi'} &=&-  \frac{(3\ell_0 + k_0^2) k_0 r}{R_1^4} + \mathcal{O}(r^2)  \; .
\eea 
In general $g_{rr}$ contains $1/r^2$ and $1/r$ singular terms, whereas $g_{r \psi'}$ contains $1/r$ singular terms.  Demanding that the $1/r$ term in $g_{r\psi'}$ and the $1/r^2$ term in $g_{rr}$ vanish, corresponds to fixing the constants
\bea
B_0 = \frac{4 k_0( 3 \ell_0+ k_0^2)A_0}{\ell_0^2 ( 8 \ell_0+ 3k_0^2)} \; , \; \; 
A_0^2 = \tfrac{1}{4} \ell_0^2 ( 8\ell_0 + 3k_0^2) .
\eea
This then gives $g_{vr} =  \pm  \frac{2}{R_2}+\mathcal{O}(r), \;  g_{r \psi'} = \mathcal{O}(1)$,  where the sign corresponds to that of $A_0$ (positive if $A_0<0$ and vice-versa). Finally, demanding that the $1/r$ term in $g_{rr}$ also vanishes, fixes $A_1$ to be a complicated constant. Then  $g_{rr} = \mathcal{O}(1)$.  Furthermore, $\chi =(1+2 \cos \theta + \mathcal{O}(r^2)) \td \phi$ and $\hat{\omega}= \mathcal{O}(r) \td \phi$ (to show the latter one needs (\ref{constr})).

It is now easily checked that the metric and its inverse are analytic at $r=0$ and therefore can be extended to a new region $r<0$. The surface $r=0$ is a degenerate Killing horizon with respect to the supersymmetric Killing field $V = \partial /\partial v$, with the upper (lower) sign corresponding to a future (past) horizon. It is also easily checked that the Maxwell field is regular on the horizon. The near-horizon geometry may be extracted by scaling $(v,r) \to (v/\epsilon, \epsilon r)$ and letting $\epsilon \to 0$~\cite{Kunduri:2013gce}. We find
\bea
\td s^2_{\text{NH} } &=& - \frac{4r^2 \td v^2}{R_1^2R_2^2} \pm  \frac{4\td v \td r }{R_2}+R_1^2( \td \theta^2 + \sin^2 \theta \td \phi^2)   \\ \nonumber  &+&\frac{R^2_2}{4} \left( \td\psi' + 2 \cos \theta \td \phi -  \frac{4(3\ell_0 + k_0^2) k_0 r \td v }{R_2^2 R_1^4} \right)^2 \\ 
F_{\text{NH}} &=& \tfrac{\sqrt{3}}{2} \td \left[ \frac{2 r \td v}{R_1^2}+ \frac{(3\ell_0+k_0^2) k_0}{2R_1^2} (\td \psi'+ 2\cos \theta \td \phi) \right], \nonumber
\eea
where we have used $k_0^2(3\ell_0 + k_0^2)^2 = R_1^4(R_1^2-R_2^2)$. This near-horizon geometry is {\it locally} isometric to that of the BMPV black hole~\cite{Breckenridge:1996is}, as guaranteed by~\cite{Reall:2002bh} (cf.~\cite{Kunduri:2014iga}). However, the period $\Delta \psi' = 4\pi$ has already been fixed by asymptotic flatness and regularity at the other centre. Therefore, cross-sections of the horizon $v=\const, r=0$ are of topology  $L(2,1) \cong \mathbb{RP}^3 \cong S^3/ \mathbb{Z}_2$, as claimed. The area of the horizon is
\be
A = 8 \pi^2 R_1^2 R_2 \; .
\ee

The above black hole solution has $U(1)^2$-rotational symmetry. The $z$-axis of the $\mathbb{R}^3$ base in the Gibbons-Hawking space corresponds to the axes where the $U(1)^2$ Killing fields vanish. We will now examine the geometry on these various axes.  Due to our choice of harmonic functions, the $z$-axis splits naturally into three intervals: $I_+ = \{ z >a_1 \} ,\; I_D = \{ 0<z<a_1 \}, \;  I_- = \{ z<0 \}$.
The semi-infinite intervals $I_\pm$ correspond to the two axes of rotation that extend out to infinity. As we will see, the finite interval $I_D$ corresponds to a non-contractible disc topology surface that ends on the horizon.

The 1-form $\chi= \pm \td \phi$ on $I_\pm$ and $\chi = 3 \td \phi$ on $I_D$. Remarkably, it can also be verified that $\hat{\omega}=0$ on the whole $z$-axis (on $I_D$ one needs to use (\ref{constr})). Thus, the geometry and Maxwell field induced on the axis are,
\bea
\td s^2_{\text{axis}} &=&  -f^2\td t^2 -\frac{\Omega_I(z) \td t \td \psi_I}{P_I(z)^2} \nonumber \\ &+& \frac{P_I(z) \td z^2 }{z^2| z-a_1|}+ \frac{Q_I(z)}{P_I(z)^2} (\td \psi_I)^2 \; , \label{axis} \\
F_{\text{axis}} &=& \tfrac{\sqrt{3}}{2} \td \left[ f\td t + \frac{R_I(z)}{P_I(z)} \td \psi_I \right]  \; , \nonumber
\eea
where $P_I, Q_I, \Omega_I, R_I$ are polynomials and  $(\psi_I, \phi_I)$ are angles, that depend on the interval. In particular, we have $(\psi_{\pm}, \phi_\pm ) =( \psi\pm \phi, \phi)$, $( \psi_D,  \phi_D)= ( \psi+ 3\phi,\phi)$ and
\bea
f = \left\{ \begin{array}{cc} \frac{z(z-2a_1)}{P_\pm(z)} , \; & z \in I_\pm \\ \frac{z(2a_1-3z)}{P_D(z)} ,\; & z \in I_D \; . \end{array} \right.
\eea
The explicit polynomials are
\bea
P_\pm (z) &=& z^2 \pm (k_1^2 +(k_0+k_1)^2 + \ell_0 \mp 2a_1)z \nonumber \\ &\mp & a_1(2\ell_0 + k_0^2)   \;, \\
P_D(z) &=& -3z^2 + (2a_1 - 3\ell_0 - k_0^2 + 2k_0 k_1 + 2k_1^2)z  \nonumber \\ &+& a_1(2\ell_0 + k_0^2) \; ,  \nonumber
\eea
whereas the $Q_I$ are quintics such that $Q_+\sim a_1^{-2}(z-a_1)P_+(a_1)^3, \; Q_D \sim a_1^{-2}(a_1-z) P_D(a_1)^3$ and $\Omega_+ = \mathcal{O}(z-a_1), \;  \Omega_D = \mathcal{O}(a_1-z)$, as $z\to a_1$.

In order for the axis geometry to be a smooth Lorentzian metric we require $P_I>0$ and $Q_I>0$ on each of their corresponding intervals.  Thus on $I_+$ we must have $P_+(z)>0$, which in fact is equivalent to $P_+(a_1)>0$ and $P_+'(a_1) > 0$ (since  $R_1^2>0$, $P_+$ has positive discriminant). Explicitly, these inequalities read
\bea
&&2k_0 k_1 + 2 k_1^2 - a_1 - \ell_0 >0 \; , \label{ineq1}\\
&& (k_0+k_1)^2+ k_1^2 + \ell_0 >  0  \; . \label{ineq2}
\eea
It is easily seen that these conditions also guarantee that $P_D(z)>0, P_-(z)>0$ on their respective intervals, since $P_D(0)= a_1 R_1^2>0, \; P_D(a_1)=P_+(a_1)>0$ and $P_-(0)= a_1R_1^2>0\; , -P'_-(0)=P_+'(a_1)+2a_1>0$.  Furthermore, we have verified numerically that in the domain (\ref{ineq1}) and (\ref{ineq2}) the polynomials $Q_\pm, Q_D$ are positive on $I_\pm, I_D$, so this places no further constraints.  Observe that $P_I>0$ also guarantees the Maxwell field is smooth.

Now, on $I_+$ the Killing field $v_+=\partial_{\phi_+}=\partial_\phi -\partial_\psi$ vanishes, whereas $\partial_{\psi_+}=\partial_\psi$ is non-vanishing everywhere and degenerates smoothly at the endpoint $z=a_1$ (one can check the conical singularity at $z=a_1$ in (\ref{axis}) is absent since $\Delta \psi_+ = 4\pi$). Next, on $I_D$ the Killing field $v_D=\partial_{\phi_D} =\partial_\phi- 3 \partial_\psi $ vanishes, whereas $\partial_{\psi_D}=\partial_\psi$ is non-vanishing everywhere and vanishes at the endpoint $z=a_1$ smoothly (again since $\Delta\psi_D=4\pi$ the conical singularity is absent). On the other hand $\partial_{\psi_D}$ does not vanish at the endpoint $z\to 0$ which ends on the horizon, so the finite interval $I_D$ is a disc topology surface $D$.  On the final interval $I_-$ the Killing field $v_-=\partial_{\phi^-}=\partial_\phi +\partial_\psi$ vanishes, whereas $\partial_{\psi_-}= \partial_\psi$ is non-vanishing everywhere including on the horizon $z\to 0$. It is worth noting that in the $2\pi$-normalised basis $( \partial_{\phi_+}, 2\partial_{\psi_+})$, the vanishing Killing fields on $I_+$ and $I_D$ are $v_+=(1,0)$ and $v_D=(1, -1)$ respectively, so the compatibility condition for adjacent intervals is satisfied~\cite{Hollands:2007aj},
\be
\det \left( v_D^T \; v_+^T \right) = 1  \; .
\ee

Finally, observe that on the axis, $f=0$ at $z=2a_1$ and $z= \tfrac{2}{3} a_1$, so the supersymmetric Killing field is null on these circles. In fact,  $P_+(2a_1) = a_1 (k_0+2k_1)^2$ and $P_D(\tfrac{2}{3} a_1)= \tfrac{1}{3}a_1(k_0+2k_1)^2$, so we must have $k_0 + 2 k_1 \neq 0$. It can be shown this implies that $\Omega_+(2a_1) \neq 0$ and $\Omega_D(\tfrac{2}{3} a_1) \neq 0$, which ensures the metric on the axis (\ref{axis}) is smooth and invertible even where $f=0$.  It is worth emphasising this is guaranteed by our above conditions. To see this, suppose $k_0=-2k_1$, so then (\ref{constr}) may be solved to get $\ell_0 = -\tfrac{4}{3} k_1^2$; in this case (\ref{ineq1}) is violated so we deduce $k_0 \neq - 2k_1$.  To summarise, we have shown that the metric on the whole $z$-axis is smooth and invertible if and only if $R_1^2>0$, (\ref{ineq1}) and (\ref{ineq2}) are satisfied.

We now address regularity and causality in the domain of outer communication $r>0$. It is easy to prove that $R_1^2>0$ and (\ref{ineq1}) imply that $K^2+HL>0$ away from the centres, ensuring $f$ is smooth everywhere. Remarkably, this also guarantees that the full spacetime metric is smooth and invertible, and the gauge field is smooth, everywhere away from the centres (even where $H=0$).  We also require stable causality with respect to the time function $t$, thus,
\be
g^{tt}= - f^{-2}+fH \omega_\psi^2 + fH^{-1} \hat{\omega}_i \hat{\omega}_i <0  \; .  \label{gtt}
\ee
Asymptotically $r\to \infty$, it is clear that this is satisfied since $g^{tt} \to -1$. Also, $g^{tt} \sim - \tfrac{1}{4}R^2_1 R^2_2 r^{-2}$ as $r \to 0$, so the solution is stably causal near the horizon. On the axes of symmetry the condition reduces to $-f^{-2}+ fH \omega_\psi^2<0$ and hence away from the centre $z=a_1$ it is equivalent to positivity of the polynomials $Q_D, Q_\pm$ discussed above.  Away from the axis we have performed extensive numerical checks and found no violation of (\ref{gtt}), provided that (\ref{regularhorizon}), (\ref{ineq1}) and (\ref{ineq2}) are satisfied. 
Therefore, we believe our solution is stably causal if and only if (\ref{regularhorizon}), (\ref{ineq1}) and (\ref{ineq2}) are satisfied.  This ensures there are no closed timelike curves in the domain of outer communication.

We will now briefly discuss some of the physical properties of our black lens solution. We find the Maxwell charge and Komar angular momenta are
\bea
Q &=& 2\pi \sqrt{3} ( \ell_0+k_1^2+ (k_0+k_1)^2) \; , \nonumber \\
J_\psi &=& \pi [ \tfrac{1}{2} k_1^3 +(k_0+k_1) ( (k_0+k_1)^2 + \tfrac{3}{2}(\ell_0+k_1^2)) ]  \; ,\nonumber \\
J_\phi &=& \tfrac{3}{2} \pi  a_1 \left( k_0  + 2 k_1 \right)  \; .
\eea
The mass is given by the BPS relation $M= \tfrac{\sqrt{3}}{2} Q$.  Our solution also carries a magnetic flux through the disc topology surface $D$ discussed above:
\be
q[D] = \frac{1}{4\pi} \int_D F = \tfrac{\sqrt{3}}{4} (k_0+2k_1)  \; .
\ee
Since our solution is a 3-parameter family there must be one constraint between these four physical quantities.  As for any BPS black hole, the surface gravity and angular velocity must vanish and the electric potential $\Phi_H = \tfrac{\sqrt{3}}{2}$. Furthermore, the electric flux $\mathcal{Q}[D]$ which appears in the first law of black hole mechanics~\cite{Kunduri:2013vka} also vanishes~\cite{Kunduri:2014iga}, so the Smarr relation and first law reduce to the BPS bound.

The magnetic flux $q[D]$ for our solution is necessarily non-vanishing, since, as shown above, smoothness of the axes of rotation requires  $k_0\neq -2 k_1$. One might be tempted to interpret the magnetic flux as `supporting'  the black lens, since the disc $D$ is required for a lens space horizon topology. However, this need not be the case. Black rings also possess a disc topology region ending on the horizon, which shows that rotation may be sufficient for supporting non-trivial topology.

One might wonder if our solution may possess the same conserved charges as the BMPV black hole. Equal angular momenta with respect to the orthogonal $U(1)^2$ Killing field at infinity, requires $J_\phi=0$ or $J_\psi=0$.  In fact $J_\phi \neq 0$ since as shown above $k_0 \neq -2k_1$. It also turns out the solution with $J_\psi=0$ is not compatible with our regularity constraints, although this is less straightforward to show. Hence there are no regular black lenses in our family of solutions, with the same charges as BMPV. On the other hand, the supersymmetric black ring possess non-equal angular momenta~\cite{Elvang:2004rt}, so we may expect there are black lenses with the same conserved charges.

In conclusion, the black lens we have presented, together with the recently found spherical black hole with an exterior 2-cycle~\cite{Kunduri:2014iga}, demonstrate that black hole uniqueness in five-dimensions is violated much more drastically than previously thought, even for supersymmetric black holes. It would be interesting to explore the implications of this for the microscopic entropy calculations in  string theory.  We also expect non-extremal versions of our solutions to exist. In particular, we do not expect magnetic flux is required to support lens space topology, so a regular vacuum black lens may also exist. 

 {\bf Acknowledgements}. HKK is supported by an NSERC Discovery Grant.  JL is supported by an EPSRC Career Acceleration Fellowship.

\end{document}